# Sub-optical resolution of single spins using magnetic resonance imaging at room temperature in diamond

Chang Shin[1,3], Changdong Kim[1], Roman Kolesov[2], Gopalakrishnan Balasubramanian[2], Fedor Jelezko[2], Jörg Wrachtrup[2] and Philip R. Hemmer[1,*]

[1]Department of Electrical and Computer Engineering, Texas A&M University, College Station, TX 77843, USA
[2]Physikalisches Institut, Universität Stuttgart, Pfaffenwaldring 57, D-70550 Stuttgart, Germany
[3]National Biomedical Center for Advanced ESR Technology, Department of Chemistry and Chemical Biology, Cornell University, Ithaca, NY 14853, USA

(Dated: January 25, 2010)

There has been much recent interest in extending the technique of magnetic resonance imaging (MRI) down to the level of single spins with sub-optical wavelength resolution. However, the signal to noise ratio for images of individual spins is usually low and this necessitates long acquisition times and low temperatures to achieve high resolution. An exception to this is the nitrogen-vacancy (NV) color center in diamond whose spin state can be detected optically at room temperature. Here we apply MRI to magnetically equivalent NV spins in order to resolve them with resolution well below the optical wavelength of the readout light. In addition, using a microwave version of MRI we achieved a resolution that is 1/270 size of the coplanar striplines, which define the effective wavelength of the microwaves that were used to excite the transition. This technique can eventually be extended to imaging of large numbers of NVs in a confocal spot and possibly to image nearby dark spins via their mutual magnetic interaction with the NV spin.

## 1. Introduction

Magnetic resonance spectroscopy has historically served as one of the fundamental methods used in understanding spin physics. In the last 25 years, it has also been used extensively for medical imaging applications where it is called magnetic resonance imaging (MRI). For this application, a magnetic field gradient is applied to the spin sample, so that the resulting magnetic resonance spectrum encodes spatial information, which is later de-convoluted to form the image. Recently, researchers have begun to push MRI down to the level of single spins where it is envisioned that individual molecules might eventually be imaged and tracked in living cells. MRI is considered to be a promising technology for subwavelength imaging because its resolution is usually considered to be independent of the excitation wavelength, being determined instead by the strength of the magnetic field gradient that can be applied. In addition to imaging, the real time monitoring and/or manipulation of single spins (or a few spins) are/is expected to significantly enhance the understanding of spin dynamics by eliminating ensemble averaging effects.

Conventional NMR or ESR techniques implemented with magnetic induction detection detect $\sim10^{12}$ or $10^7$ spins, respectively, with a resolution of a few micrometers in an hour. Recently it was reported that MRFM can probe a single electron spin [1] and image the proton density in the individual tobacco mosaic virus with better than 10 nm resolution [2], but requires ultra high vacuum and cryogenic temperatures, which may not be desirable for many applications.

In contrast, optically detected ESR (ODESR) is known to be capable of single spin detection. Nitrogen-vacancy (NV) diamond in particular has demonstrated single spin detection under ambient conditions. The NV center is unique among ODESR systems in that it has very long electron spin coherence times ($T_2\sim1.8$ms or longer in $^{12}$C enriched diamond) even at room temperature [3]. Its spin state can also be optically initialized and read out with high contrast using either laser or incoherent light. Recently, using a magnetic tip to produce a large field gradient, NV diamond has achieved 5 nm resolution under ambient conditions [4]. In this work, the nanocrystals containing NVs were also investigated as a potential spin label for biological cells. In these first experiments, the high resolution was achieved by using a technique based on curve fitting to accurately find the center frequency of the spectral line. This technique is applicable when the chosen NVs are magnetically in-equivalent, and do not require the magnetic gradient to be spectrally resolved.

In this paper we extend MRI to magnetically equivalent NVs. This opens the possibility for scaling to large numbers of NVs within the same focal spot. To achieve the necessary field gradients, we use micro-electromagnets similar to those used to trap and manipulate atoms in vacuum at low temperature ($\sim20$ K) [5]. Further, we image with both dc and microwave magnetic gradients and show that the resolution is similar. In the case of microwave fields, the co-planar striplines used to produce the field gradient define an effective wavelength of the microwave field, which is much smaller than the free space wavelength. The resolution of MRI-enhanced microwave imaging is orders of magnitude better than even the confined wavelength. Once these ESR techniques are integrated into a confocal microscope they promise to become a very powerful diagnostic for biological processes in living cells.

---

* Corresponding author. Tel.: +1-979-845-8932; fax: +1-979-845-6259.
*E-mail address*: prhemmer@ece.tamu.edu.

## 2. Experimental setup and spin sample characterization

A home-built laser scanning confocal fluorescence microscope (LSCFM) was assembled on an optical table using an oil immersion objective (NA=1.4) and a pinhole of 50 μm in diameter. To achieve single molecule sensitivity, photon counting avalanche photodiodes (APD) were used as detectors. To produce strong magnetic gradients that could be rapidly pulsed, gradient wires ~10 μm widths were fabricated on a quartz coverslip of 150 μm thickness. These were integrated with microwave (MW) coplanar strip-lines, also of ~10 μm width, that were used to excite the spin coherences, and also to produce the microwave field gradients used in the MW imaging demonstration (details are found elsewhere [6]).

The NV diamond sample was a type IIa single crystal, bulk diamond with a low enough concentration of NV defects to give some resolvable spots, containing single NVs. This sample was placed in contact with the fabricated microelectronic structure and imaged through the quartz coverslip.

To achieve maximum spin contrast all experiments were carried out in pulse mode so that the optical and microwave fields were never present at the same time. This avoids any unwanted spin re-polarization or mixing by the optical fields that would broaden the spin resonances and reduce imaging contrast. Even the usual CW electron spin resonance (ESR) measurements were done in a pulsed mode. Other measurement techniques that were used include electron spin echo envelop modulation (ESEEM), and direct Rabi nutation experiments.

A typical fluorescence image acquired by the confocal microscope is shown in Fig. 1(a). Here NV defects in the diamond sample appear as bright spots located between the microstructure wires. To determine the number of NV centers inside a single confocal microscope spot, second–order photon antibunching experiments were performed using the Hanbury-Brown-Twiss setup with two avalanche photo detectors (APDs) [7,8,9]. In this image more than 40 NV sites with antibunching dips between 0.5 and 0.9 were found. Representative antibunching traces are shown in Fig. 1(b) for the cases of a single NV, a pair of NVs and more than 2 NVs.

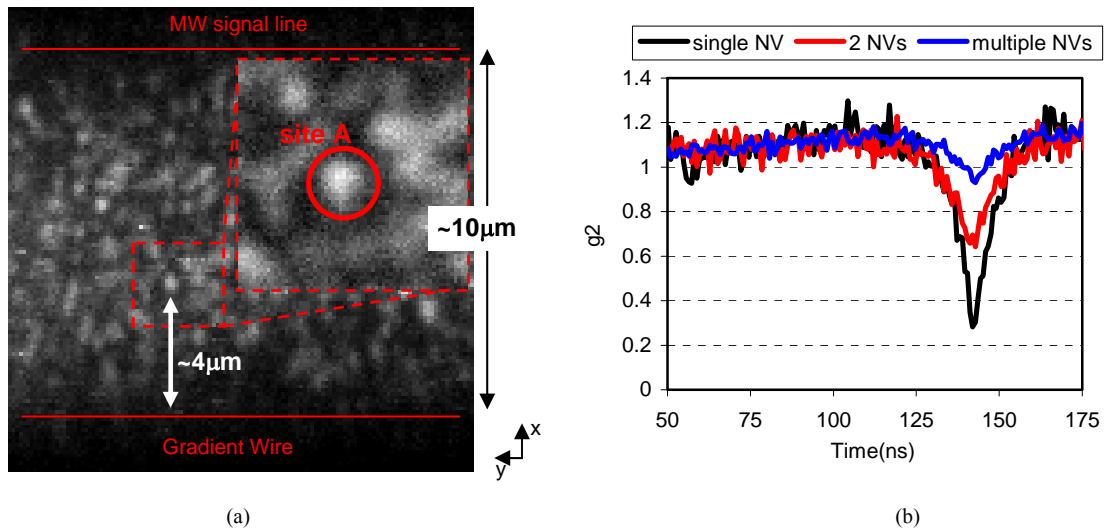

(a)                                                                                    (b)

Fig. 1. (a) Fluorescence image of NV defects between a microwave signal line (~10μm wide) and one of the dc gradient wires (~10μm). Site 'A' was located at x=9.6μm, and z=-2.6μm from the center (not the edge) of the gradient wire, where the x-direction is perpendicular to the wires. (b). Representative photon anti-bunching data for a single NV defect, 2 NV defects (on 'site B'), and multiple NV defects (on 'site A'). Optical pumping power is less than 1mW.

Here, it should be noted that the antibunching traces in Fig. 1(b) include background subtraction where the background was obtained by accumulating counts from a dark region in the image for the same acquisition time. This background is due to fluorescence from the diamond surface, the quartz coverslip, and the immersion oil in between, as well as dark counts of the avalanche photodiode (APD). Without background subtraction the antibunching dip for a single NV defect was ~0.3. From these 40 sites, the one named as 'site A' in Fig. 1(a) near the DC gradient wire was selected to demonstrate MRI imaging with a DC magnetic field gradient. Another NV site named as 'site B' (not shown) closer to the microwave (MW) signal wire was selected to demonstrate MRI using Rabi nutations with a MW field gradient.

In NV defects the dominant contributions to the spin Hamiltonian are the zero field splitting and the electron Zeeman interaction. In this approximation, the spin Hamiltonian can be simplified as $\mathbf{H}= \mathbf{SDS}+g_e\beta_e\mathbf{BS}$, where $\mathbf{H}$ is the Hamiltonian for the total energy of the quantum system, $\mathbf{D}$ is the zero field splitting tensor, $\mathbf{B}$ is the external magnetic field, $\mathbf{S}$ is the electron spin vector, and $g_e\beta_e$ =2.8 MHz/Gauss. According to this Hamiltonian, NV defects whose quantization axis orientations are not identical can be spectrally resolved by using the dependence of the energy splitting on the angle between NV orientation and the magnetic field direction.

In order to verify the orientations of the NV defects in 'site A' or 'site B', ESR spectra were first obtained, while a uniform external DC magnetic field of ~18 Gauss or ~110 Gauss, respectively, is applied along the [111] crystal direction, which is perpendicular to the diamond surface e.g. z direction (data are shown elsewhere [10]). Then additional ESR experiments with moderate uniform magnetic fields applied at various angles in x-z or y-z plane confirmed that there are only two distinct orientations for the NVs in 'site A' and only one in 'site B' (data is shown elsewhere [6]). From these ESR spectra, it was concluded that most of the NV defects in 'site A' and all of those in 'site B' are oriented along the z axis.

In the present configuration where a single wire provides the magnetic field gradient, it will be superimposed on a non-zero magnetic bias field due to the current in the wire. In general, this bias field does not have to be parallel to the z axis, along which the static magnetic field is applied. Therefore it is necessary to determine the angle of the bias field relative to the NV quantization axes. This can be done by applying different currents to the DC (or MW) gradient wire, measuring the ESR spectral shifts, and then using the above Hamiltonian to determine the angle by fitting the data. From the ESR transition frequencies measured at various currents along the dc (or MW) gradient wire, the angle for the NVs of interest in site A or B is estimated to be $51.3^{o}$ or $49.6^{o}$ (data is shown elsewhere [10]).

## 3. One dimensional single spin imaging

### 3.1. Quasi CW-ESR experiments

As current is applied to the gradient wire, the ESR lines shift according to:

$$\Delta \omega_{\pm 1} = \pm \frac{2 g_{NV} \mu_B}{\hbar} |\Delta B| \cos \theta \qquad (1)$$

for the $m_s = +/- 1$ spin sublevels, respectively, where $\theta$ is the angle between the NV and the bias field from the gradient wire. Here it is assumed that the bias field is larger than the uniform magnetic field applied along the z-axis, which is true in the experiment for all but the smallest gradient currents. For MRI demonstrations with CW ESR or ESEEM experiments, the strongest ESR line was used, i.e. the transition line at 2.92 GHz in ESR spectra of NVs in site A.

Since the strength of the magnetic field depends on the distance from the wire, NVs defects located at different distances from the wire will have different spin transition frequencies. Thus for the multiple NVs in site A, the ESR lines broaden as the current in the gradient wire increases and will eventually begin to split. This is shown in Fig. 2 for a large current of 1.6 A in the gradient wire. In this case, it is clear that the transition line is no longer a single Lorentzian peak, but is best fit by a superposition of three Lorentzian peaks with splittings of 8.11 MHz and 4.78 MHz.

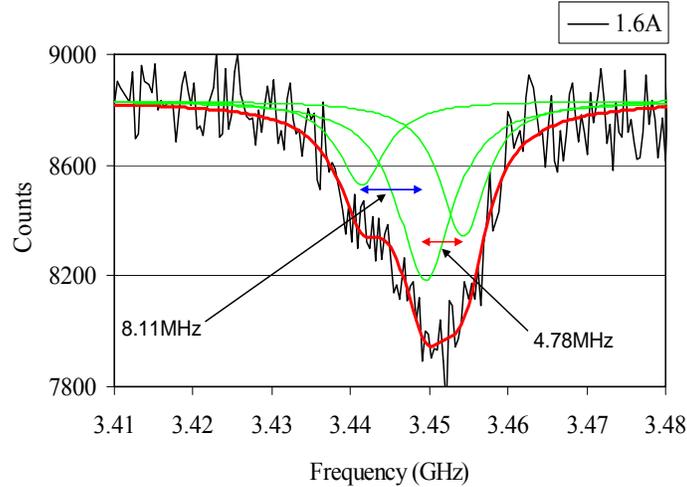

Fig. 2. ESR spectrum of NV defects in site A with a current of 1.6A applied to the gradient wire and a uniform dc magnetic field of 18 Gauss along z axis. Solid lines are Lorentzian fits. The frequency difference between neighboring Lorentzians is 8.11MHz and 4.78MHz, respectively.

When a smaller current of 1.4 A applied to the gradient wire, the frequency differences reduce to 6.96 MHz and 4.23 MHz, respectively, and to 6.16 MHz and 3.61 MHz for a current of 1.2 A, which is consistent with the splittings being caused by the gradient field . At the highest current, the spectral lines should have been better resolved, since linewidth in the absence of the gradient is only ~5 MHz. However, there is an unknown source of spectral broadening at high gradient currents, since the fitted Lorentzians in Fig. 2 each have widths of ~7.5 and 8.5 MHz.

Such broadening can be caused by a variety of effects, for example motion of the gradient wire relative to the diamond crystal due to local heating which might give rise to a frequency chirp, or ringing of the gradient current caused by switching and impedance mismatch. To overcome these and other possible broadening effects spin echoes (ESEEM) were used next.

### 3.2. ESEEM Experiments

In NV diamond, spin echo experiments are performed with the modified Hahn echo pulse sequences ($\pi/2$-$\tau_1$-$\pi$-$\tau_2$-$\pi/2$). Since these are not sensitive to static magnetic fields including any gradient fields, the gradient field cannot be CW but must be pulsed, for example switched on only during the $\tau_1$ time.

In order to carry out the ESEEM enhanced MRI, the first and second free evolution times of the echo pulse sequence are set equal so that the echo would have a maximum amplitude in the absence of the gradient field. Then the magnetic gradient is pulsed on for a short time, only during the first free evolution time, causing a change in the echo amplitude. The echo sequence is repeated many times, with the gradient field being turned on for a progressively longer duration each time. Then, plotting the echo amplitude as a function of gradient pulse duration gives a sinusoidal signal that shows how the phase difference between the microwave field and the spin polarization accumulates as a function of time. For a single NV, only a simple sine wave would result. However, for two NVs with slightly different positions, two sine waves are produced and a beating would be observed.

Again, under the uniform magnetic field of ~18 Gauss applied along the z axis, the 2.92 GHz frequency component of the NV defects in site A is used for this ESEEM experiment. A MW $\pi$-pulse of 50 ns duration is chosen to be fast enough to allow relatively large gradient fields to be used but narrowband enough to prevent significant off resonance excitation of the magnetically in-equivalent NVs.

First, a small current of 0.04 A is applied to the gradient wire. The resulting variation of echo amplitude with gradient pulse "on-time" is plotted in Fig. 3(a). This relatively small current produced a magnetic field of ~ 5.6 Gauss, which shifts the $m_s$ = +1 sublevel of the NV electron spin by ~9.8 MHz and gives the observed echo oscillation or modulation at that frequency. Only one oscillation frequency is seen, even though there are multiple NVs present, because the phase accumulated by each NV is very small at low current.

When the current is increased to 0.2 A, a higher magnetic field of ~27 Gauss is produced resulting in a higher frequency oscillation of the echo at ~47.3 MHz as shown in Fig. 3(b). At this current the gradient field was large enough to produce a beating in the echo modulation due to interference of signals produced by the different NVs. As seen, there are two major frequency components resulting in a strong modulation with a clearly defined node at 768 ns. Once the first node is observed in the echo modulation signal, the magnetically equivalent NVs can be considered to be spectrally and therefore resolved.

In order to determine the maximum resolution of the system the experiment was repeated with higher currents up to 0.8 A. As current increased, the difference between the neighboring frequency components increases, and more beats become visible as shown in Fig. 4. A third frequency component also becomes more evident at higher currents.

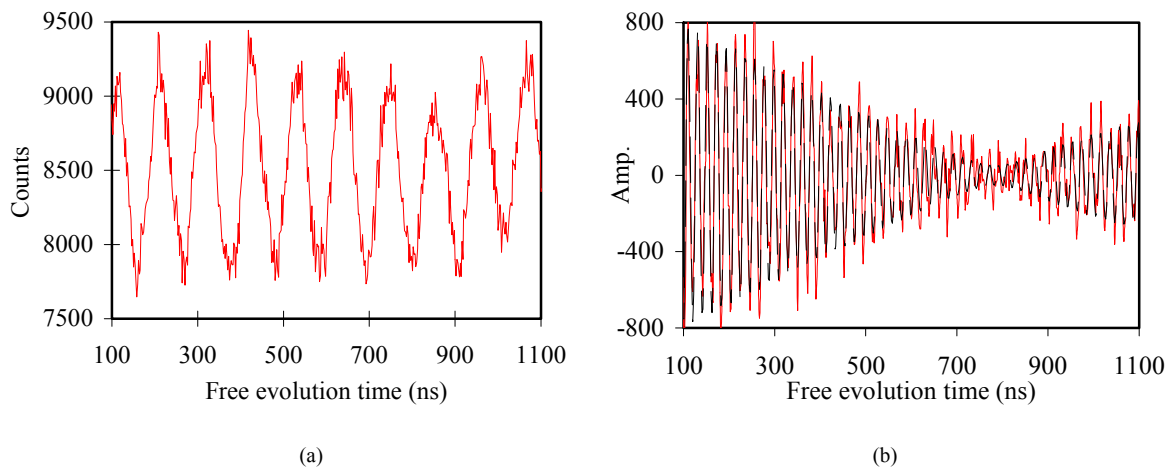

(a)                          (b)

Fig. 3. ESEEM experiments with a magnetic field produced by a current applied to the gradient wire. (a) Modulation frequency of 9.78MHz, but no beating is observed at current of 0.04A, (b) At higher current of 0.2A a beating is observed and the major frequency components are 47.99MHz and 47.34MHz. The first node occurs at about 768ns (measured data are in red, and the fitted sine wave pattern is a black, dashed line).

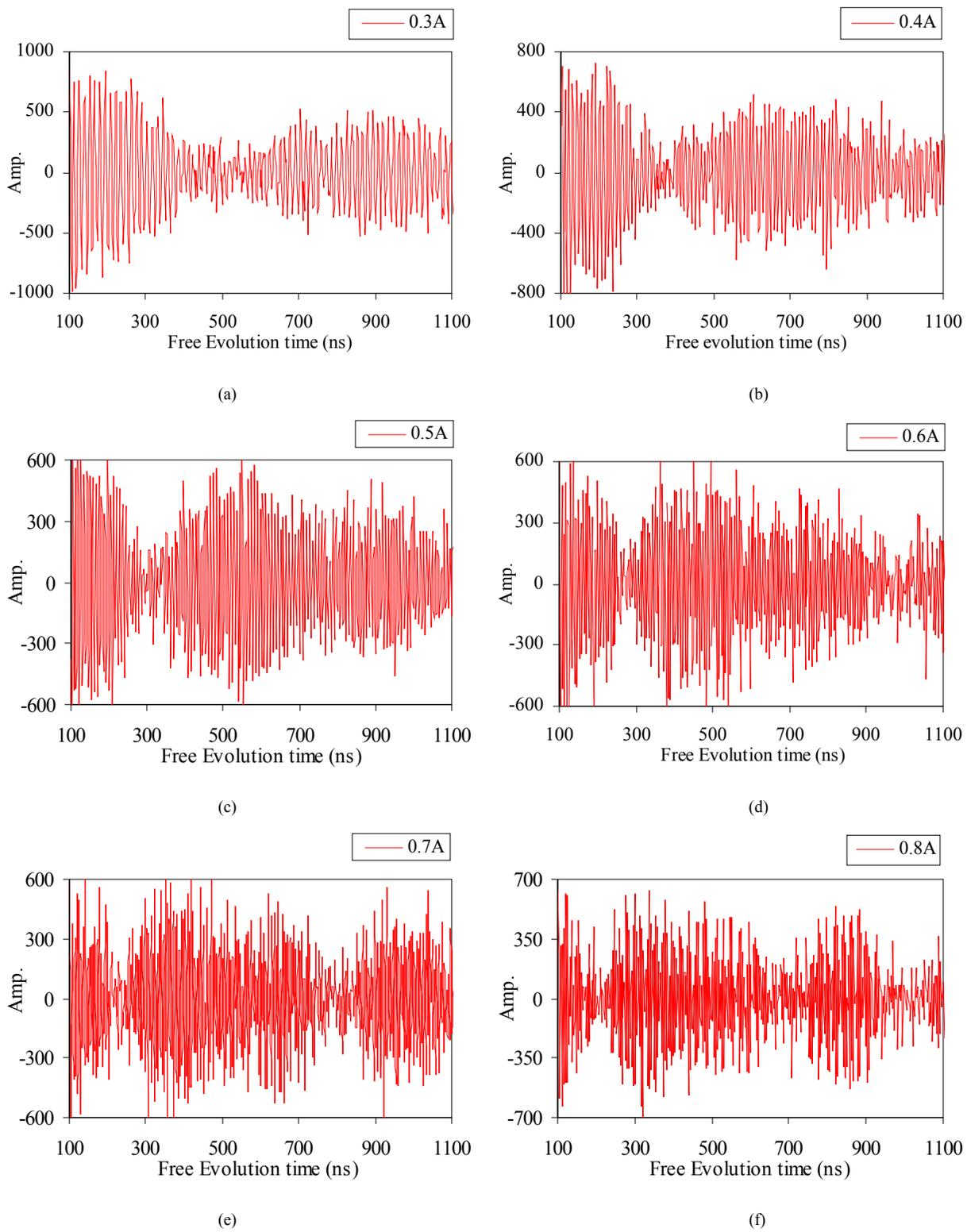

Fig. 4. ESEEM signals at various dc currents applied to the gradient wire.

In order to identify broadening mechanism such as frequency chirps, a short-time Fourier transform (STFT) was applied to all the data, and a noticeable frequency chirp shorter than 400ns was observed. Frequency spectra of ESEEM signal, which were obtained after employing appropriate window functions to compensate for this chirp showed clearly resolved three frequency components (data is shown elsewhere [10]).

To determine the positions of the three NVs in site A, the echo modulation frequencies (derived from curve fits with 3 frequencies) are plotted as a function of the gradient current in Fig. 5(a) and the difference frequencies between the NV transitions are plotted vs. gradient current in Fig. 5(b). The ESR frequencies determined from the earlier CW experiments are also added to the plot. For the purpose of data analysis, the highest frequency component of the echo modulation in 'site A' is named 'NV1', the next highest is 'NV2', and the lowest frequency is 'NV3'.

As seen in Fig. 5(a) both the frequency differences and the absolute frequencies of the NV transitions increase with the gradient current. The absolute frequency increase is due to the bias field component and shows a small quadratic dependence due to the angle between the NV axis and the bias field direction. As shown in Fig. 5 (a) and (b), there is good agreement between the CW-ESR data and the ESEEM data.

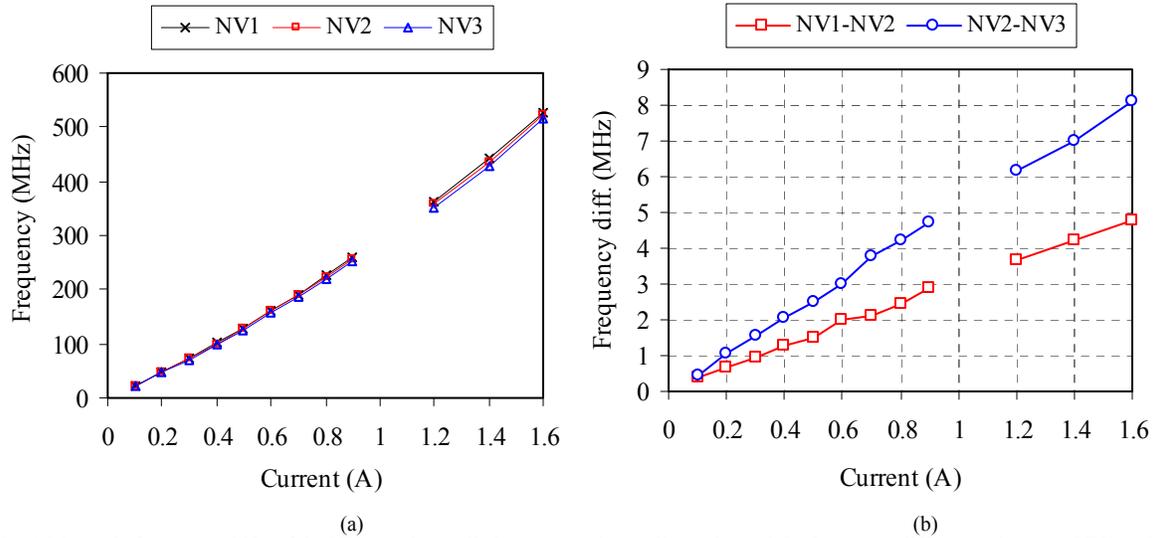

(a)                                                                                    (b)

Fig. 5. (a) Magnetic frequency shifts of the NVs vs. the applied current to the gradient wire and (b) frequency differences between NV1 and NV2 (squares) and NV2 and NV3 (circles).

To convert the frequency differences into distances, the numerical value of the magnetic field gradient is required. This is estimated using the measured distance of ~ 9.95 μm from the center of the gradient wire to NV 'site A,' as discussed in Fig. 1. At this relatively large distance, we can use the thin wire approximation for the induced magnetic field around the gradient wire. This approximation will be justified later.

The magnetic field outside a wire carrying a current $I$ can be described by the Biot-Savart law: [11]

$$\mathbf{B}_\theta\left(r\right) = a_\theta \frac{\mu_o \mu_r I}{2\pi r}\left(Tesla\right) \qquad (2)$$

Since the magnetic field from the gradient wire is oriented at about 51.3° from the z axis of the dominant NVs in 'site A', its z component is:

$$\left|\mathbf{B}_z\left(r\right)\right| = cos\theta \frac{0.2 \cdot I\left(A\right)}{r\left(\mu m\right)}\left(Tesla\right) = 1250.5\frac{I\left(A\right)}{r\left(\mu m\right)}\left(Gauss\right). \qquad (3)$$

The small quadratic dependence of the frequency vs gradient wire current, seen in Fig. 5 (a), can now be ignored as we are interested in difference frequencies and these do not show significant quadratic dependance (see Fig. 5 (b)) to first order. Thus the frequency difference for NVs located at distances $r_1$ and $r_2$ from the wire center are:

$$\Delta E = \left(\Delta E_{0\leftrightarrow+1}\left(r_1\right) - \Delta E_{0\leftrightarrow+1}\left(r_2\right)\right) = \left(\left|B_z\left(r_1\right)\right| - \left|B_z\left(r_2\right)\right|\right) \cdot \left(2.8MHz/Gauss\right) \qquad (4)$$

By fitting the experimental data of Fig. 5(b) to the above Eqs. (3) and (4), the radial distances of the magnetically equivalent NVs in 'site A' from the wire center are roughly determined as: $r_{NV1}$=9.862 μm, $r_{NV2}$=9.946 μm, $r_{NV3}$=10.092 μm. So the separation between NV1 and NV2 is about 84 nm, and the distance between NV2 and NV3 is about 146 nm. This is consistent with the approximate confocal spot size of 300 nm which is almost enough to resolve NV3. A more accurate calibration check will be performed later in the MW version of MRI.

Here it should be noted that the resolution is higher than the distance between NVs since the NV1 and NV2 are first resolved at a current of 0.2 A. Taking the maximum ESEEM current of 0.8 A this gives a resolution or 21 nm. It should be emphasized that this is only limited by the memory of the pulse generator. The NV coherence lifetime is ~100 μs but the generator memory limitation stops the acquisition time at ~1 μs, and so with a better pulse generator the resolution would be ~100 times better, or 0.2 nm, in principle.

### 3.3. Rabi nutation experiments

So far conventional MRI has been explored. However the microwave field in the vicinity of the coplanar striplines also has a strong gradient and it should be possible to image with this instead of the pulsed DC magnetic gradient. To see how this can be done, recall that when a quantum system, like the electron spin of an NV defect, is driven by a strong MW magnetic field resonant with a transition from the lower ($\psi_0$) to the upper state ($\psi_1$), the population oscillates between the two states at the Rabi frequency ($\Omega$) given by,

$$\Omega = \frac{g\mu_B}{h}\left\langle \psi_1 \middle| S \times B_1 \middle| \psi_0 \right\rangle. \tag{5}$$

If the NV defects are in a spatially non-uniform MW magnetic field ($B_1$), then the Rabi nutation frequency of each NV defect would be different depending on location. To demonstrate this imaging modality the NV defect 'site B' (not shown) was selected, because it is located relatively close to the MW signal line, about 8.212 μm from its center. The optical pulses for the preparation and detection are the same as those in the modified CW ESR experiments, but there is no DC gradient current, and instead of π pulses, the microwave field is applied for varying lengths of time, ranging from 2 ns to 1024 ns with an increment of 2 ns.

For MRI demonstration with Rabi nutation experiments, the microwave is tuned to the $m_s=0 \rightarrow m_s=-1$ transition frequency, i.e. 2.562 GHz, under uniform magnetic field of ~110 Gauss applied along the z-axis, which is enough to split the $m_s=+/-1$ transitions by a large amount to prevent unwanted off resonance spin excitation of thetransition, i.e. $m_s=0 \rightarrow m_s=+1$ from being excited by the strong microwave field.

Since the maximum number of Rabi oscillations for a fixed excitation time depends on microwave power, this power was stabilized using a home-built feedback circuit.

At a MW current of about 20 mA the Rabi nutation period is 100 ns as shown in Fig. 6 (a). Though the data shows a modulation in amplitude with time, it is not clear if this is due to decay or beating of the Rabi frequency components from the two NVs.

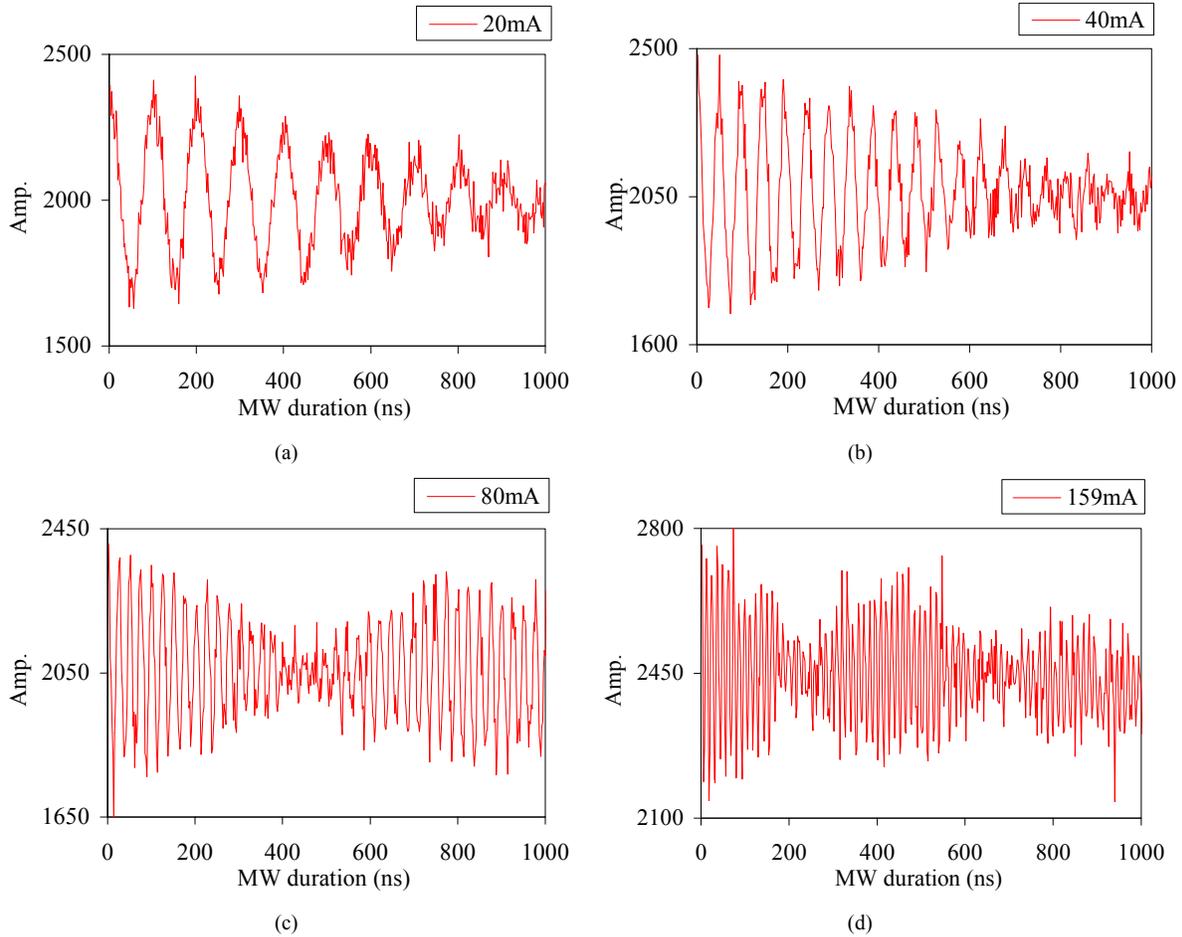

Fig. 6. Rabi nutations of electron spins in two magnetically equivalent NV defects at site B are plotted as a function of MW pulse duration. MW current is (a) 20mA, (b) 40mA, (c) 80mA, (d) 159mA, (e) 207mA and (f) 243mA

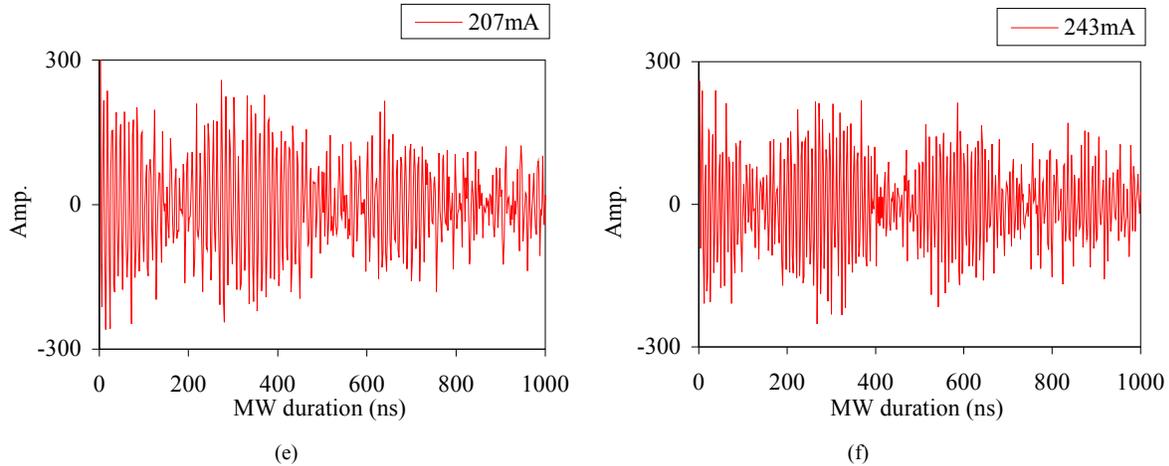

Fig. 6. Continued.

To resolve this question, the MW current is increased up to 243 mA as seen in Fig. 6 (b), (c), (d), (e) and (f). Now the nutation frequencies are faster and a beating is clearly observed, even at a MW current as small as 40 mA. Since the first beat minimum actually appears in Fig. 6 (b) for the case of 40 mA, the NVs can be considered to be just resolved at this MW current. After taking the FFT of the nutation signals, the resulting frequency spectra showed clearly resolved two peaks (data is shown elsewhere [10]).

In order to verify that this beating originates from the MW magnetic gradient instead of an experimental artifact such as nitrogen hyperfine coupling, Rabi nutation experiments are performed on a nearby single NV defect, which is located at ~9.1 μm from the MW signal wire. No beating was observed, as seen in Fig. 7. The decay of Rabi oscillations at high MW power for the single NV in Fig. 7 is likely due to residual MW power variations that are estimated at ~1%.

At a MW current of 243 mA, the nutation frequency of this single NV spin is 114.25 MHz, compared to ~127 MHz for the two NVs in nearby site B. This difference in Rabi frequencies is due to the shorter distance of site B, i.e. 8.2 μm from the MW wire center, compared to ~9.1 μm for the single NV. Since these distances are determined from the image in Fig. 1(a), they serve as an important calibration that can be used to check the validity of the approximations used in deriving Eqs. (4) and (6). In fact, by inserting these distances into Eq. (6) the resulting Rabi frequency difference is within a few percent of that actually measured.

The two Rabi frequencies obtained by fitting the data of Fig. 6 (site B) to a sum of two sine waves are plotted as a function of the MW current in Fig. 8 (a), and the beat frequency is plotted in Fig. 8 (b). As seen, the difference frequency depends linearly on the MW excitation current with an intercept at zero, except for the lowest Rabi frequency where there was not sufficient time to observe a beat minimum.

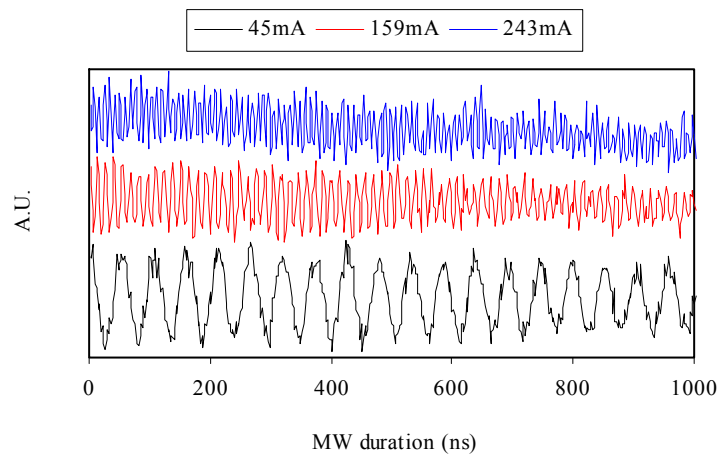

Fig. 7. Rabi nutations of a single NV spin at various MW currents.

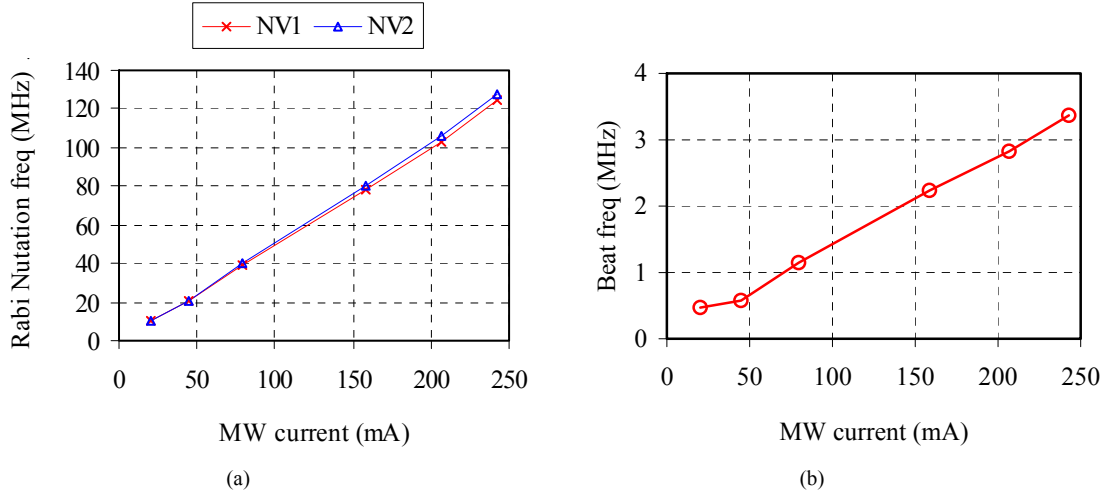

Fig. 8. (a) Rabi nutation frequencies of two NV defects at site B as a function of the applied MW. Red line is for the NV1 and blue line for the NV2. (b) Beat frequency of the two Rabi nutations is plotted as a function of the MW current.

Again the thin wire approximation is used to estimate the distance between the two magnetically equivalent NV defects in 'site B.' Although the approximation is not expected to be as good for MW currents as for DC currents, the calibration described above justifies it.

Since the DC and MW magnetic Zeeman factors are the same, Eqs. (5) and (2), can again be used to compute the distances between NVs from the Rabi frequency difference according to

$$\Delta\Omega = (\Omega_{r1} - \Omega_{r2})$$

$$= 2.8 \left( MHz / Gauss \right) \cdot \sin\theta \cdot 2000 \cdot \left( \frac{1}{r_1 \left( \mu m \right)} - \frac{1}{r_2 \left( \mu m \right)} \right) \cdot I(A) \qquad (6)$$

Fitting the experimental data shown in Fig. 8 (b) to Eq. (6) with $\theta = 49.6^{\circ}$ (determined earlier for site B), the location of the two NV defects are computed to be $r_1$=8.102 µm, and $r_2$=8.322 µm, and so the estimated distance between the two NV defects is about 220 nm. As before the resolution is higher than this separation. Since the NVs are just resolved at 40 mA but currents up to 243 mA are applied, the resolution is a factor of 6 higher than the separation or ~37 nm. This is comparable to the resolution in the pulsed DC gradient field, as predicted by Eqs. (4) and (6).

Finally, to demonstrate the feasibility of using the fabricated devices for a variety of applications requiring ultra-fast spin manipulations, Rabi nutation experiment were carried out at a microwave input power of 10 Watts. A Rabi nutation period of 1.33 ns was observed, which corresponds to an oscillation frequency of 732 MHz, or $B_1$ field of ~261 Gauss (data is shown elsewhere [6]). For these measurements nano diamond crystals were spin-coated onto the fabricated device so that it would be possible to find NVs very near the wire with the optimum orientation relative to the microwave magnetic field. To get a feel for the applications impact of such a high Rabi frequency, note that this fast Rabi frequency can be used to perform nearly $10^6$ single qubit quantum gate operations within the spin dephasing time.

## 4. Discussion

Using ESEEM techniques together with a steep magnetic field gradient near a micro-strip wire, three magnetically equivalent NV defects in 'site A' were identified and imaged in 1D. The positions determined by echo techniques agreed well with those deduced by fitting the CW-ESR lines, as shown in Fig. 5. The estimated inter distance between NV1 and NV2 is about 84 nm, and the distance between NV2 and NV3 is about 146 nm. The resolution of this MRI scheme is higher than the separation between NV1 and NV2 as these centers are resolved at a relatively low gradient current. The maximum resolution in this experiment is estimated at 21nm, limited only by the memory size of the pulse generator. In particular, the pulse generator memory limited the maximum acquisition time to ~1 µs, yet the coherence time of NVs in the sample used is > 100 µs and hence 100 times higher resolution should be possible under nearly identical experimental conditions but with a more pulse generator memory.

At the highest current of 1.6 A, the induced magnetic field gradient is about 32 Gauss/µm or 20 Gauss/µm per 1 A, which is relatively small compared to what has been seen before with similar microstructures, and therefore even higher resolution should be possible. For example, by fabricating micrometer sized devices on a substrate with higher thermal conductivity, such as diamond, much higher current (e.g. a few Amperes) can be applied to induce higher magnetic field gradients. Thus a few nanometer resolution should be easily achievable while still keeping a relatively large >10 μm spacing between wires, as desired for biological applications.

Finally, in the MW gradient imaging experiment the distance between the two NV defects is estimated to be ~220 nm. The maximum resolution is a factor of 6 smaller than this or about 37 nm. This is similar to the DC MRI owing to both having the same g-factor and the fact that similar gradient coil geometries were used for DC and MW gradient. Here, in addition to pulse generator memory, the resolution is also limited by the microwave current available, and also to the MW power fluctuations. The estimated $B_1$ field gradient at ~243 mA is about 7.3 Gauss/μm or 30 Gauss/μm per Amp current, which is larger (per Amp current) than the dc field gradient. This is due to the relatively shorter distance of site B to the MW gradient wire than site A.

## 5. Conclusion

In conclusion, optically detected ESR techniques including CW ESR, ESEEM, and Rabi nutations have shown the ability to achieve high sub-optical resolution when implemented with micrometer sized gradient wires and co-planar microwave strip-lines. The resolution is 21 nm for usual DC version of MRI and 37 nm for the microwave (MW) version. This corresponds to $\lambda^{opt}/25$ and $\lambda^{opt}/14$ respectively where $\lambda^{opt}$ is the wavelength of the 532 nm optical excitation. However, in the MW version of MRI the microwave wavelength determines a Rayleigh limit for the resolution. In this case the 37 nm corresponds to a subwavelength resolution of $\lambda^{effMW}/270$ where $\lambda^{effMW}$ (= 10 μm) is the effective Rayleigh limit for a MW confined on the 10 μm stripline. If this resolution is compared to the free space MW wavelength of $\lambda^{MW} =10.4$ cm (for the 2.88 GHz excitation) the resolution would be $\lambda^{MW}/2.8M$. But this is only the beginning, because although these resolutions are an extremely small fraction of the exciting wavelength, they still mainly limited by the memory in the pulse generator, and hence resolutions of well below 1nm should be possible in the future. As a result, MRI with NV spin labels shows great promise as a future tool to study and manipulate spins in individual NVs with nanometer-scale resolution. This can find many applications including imaging nuclei in single molecules via coupling to a nearby NV, atomic and/or molecular physics, and spin based quantum information processing at room temperature.

### Acknowledgements

We acknowledge Aleksander Wojcik, Zhijie Deng, Mughees Khan and Elizabeth Trajkov for the help of the setup, and Peter P. Borbat for the discussions on the pulse gradient driver design. This research was supported by DARPA.